\documentclass[11pt]{article}
\begin{document}
\title{{\bf{\Large Quantum Tunneling Beyond Semiclassical Approximation}}}
\author{
 {\bf {\normalsize Rabin Banerjee}$
$\thanks{E-mail: rabin@bose.res.in}},\, 
 {\bf {\normalsize Bibhas Ranjan Majhi}$
$\thanks{E-mail: bibhas@bose.res.in}}\\
 {\normalsize S.~N.~Bose National Centre for Basic Sciences,}
\\{\normalsize JD Block, Sector III, Salt Lake, Kolkata-700098, India}
\\[0.3cm]
}

\maketitle

{\bf Abstract:}\\
      Hawking radiation as tunneling by Hamilton-Jacobi method beyond semiclassical approximation is analysed. We compute all quantum corrections in the single particle action revealing that these are proportional to the usual semiclassical contribution. We show that a simple choice of the proportionality constants reproduces the one loop back reaction effect in the spacetime, found by conformal field theory methods, which modifies the Hawking temperature of the black hole. Using the law of black hole mechanics we give the corrections to the Bekenstein-Hawking area law following from the modified Hawking temperature. Some examples are explicitly worked out.\\

\section{Introduction}
         Classical general relativity gives the concept of a black hole from which nothing can escape. In 1974 Hawking \cite{Hawking1} startled the physics community by proving that black holes are not black; they radiate energy continuously. Later on in 1975, he \cite{Hawking2} showed that the radiation of the black hole perfectly matches with the black body radiation whose temperature is $T=\frac{{\cal{K}}}{2\pi}$, where ${\cal{K}}$ is the surface gravity of the black hole. His calculation was completely based on quantum field theory.

         Since the original analysis was technically very involved, several derivations of Hawking radiation were subsequently presented in the literature \cite{Hartle,Gibbons,Fulling}. None of them, however, corresponds directly to one of the heuristic pictures that visualises the source of radiation as tunneling. This picture is similar to an electron-positron pair creation in a constant electric field. The idea is that the energy of the particle changes sign as it crosses the horizon, so that a pair created just inside or outside the horizon can materialise with zero total energy, after one member of the pair has tunneled to the opposite side. In this method \cite{Wilczek,Paddy}, the particles are allowed to follow classically forbidden trajectories, by starting just behind the horizon onward to infinity. The particles then travel back in time, since the horizon is locally to the future of the external region. Thus the classical one particle action becomes complex and so the tunneling amplitude is governed by the imaginary part of this action for the outgoing particle. However, the action for the ingoing particle must be real, since classically a particle can fall behind the horizon. This is an important point of our calculations as will be seen later. The essence of tunneling based calculations is, thus, the computation of the imaginary part of the action for the process of $s$-wave emission across the horizon, which in turn is related to the Boltzmann factor for the emission at the Hawking temperature. There are two different methods to calculate the imaginary part of the action: one is by Parikh-Wilczek \cite{Wilczek} - radial null geodesic method and another is the Hamilton-Jacobi method which was first used by Srinivasan et. al. \cite{Paddy}. Later, many people \cite{Jiang} used the radial null geodesic method to find out the Hawking temperature for different spacetime metrics. Recently \cite{Chen}, tunneling of a Dirac particle through the event horizon was also studied. All these computations are, however, confined to the semiclassical approximation only. The issue of quantum corrections is generally not discussed. Inspite of some sporadic attemps \cite{Medved1,Medved2} a systemetic, thorough and complete analysis is lacking. In our previous work \cite{Majhi} we found out the corrections to the temperature and entropy by including the effects of back reaction knowing the modified surface gravity of the black hole due to one loop back reaction for the Schwarzschild case by radial null geodesic method. As an extension we \cite{Bibhas} also applied this method for a noncommutative Schwarzschild metric. Recently, a problem in this approach has been discussed in \cite{Borun,Singleton,Akhmedov,Douglas,Pilling} which corresponds to a factor two ambiguity in the original Hawking temperature.

In this paper, we formulate the Hamilton-Jacobi method beyond the semiclassical approximation by considering all the terms in the expansion of the one particle action. We show that the higher order terms are proportional to the semiclassical contribution. By dimensional argument the form of these proportionality constants, upto some dimensionless parameters, are determined. In particular, for Schwarzschild and Anti-de Sitter Schwarzschild spacetimes, these are given by the inverse powers of the square of the mass of the black hole, because in these cases the only macroscopic parameter is mass. By using the principle of ``detailed balance'' the Hawking temperature is identified. It involves a correction to the usual semiclassical approximation. Incidentally, the form of the semiclassical Hawking temperature found here is general enough to consider examples of metrics that need not be spherically symmetric. We also show that this temperature reduces to the standard form for spherically symmetric metrics. The modified Hawking temperature which includes corrections to the semiclassical structure is also derived. We show that for an appropriate choice of the dimensionless parameters, it is possible to reproduce the one loop back reaction effects \cite{York,Lousto} in the spacetime or the quantum corrections, obtained by conformal field theory techniques, due to trace anomaly. Also, we analyse the two loop corrections in the surface gravity. The Hamilton-Jacobi method is then discussed in other coordinates like Painleve. Here we show how the standard form of the Hawking temperature can be recovered. There are no factor two ambiguities which are usually reported in dealing with tunneling methods \cite{Singleton,Pilling}. Using the law of black hole mechanics the corrections to the Bekenstein-Hawking area law for Schwarzschild and  AdS Schwarzschild black holes are explicitly calculated. Interestingly the leading order correction to the Bekenstein-Hawking entropy is the famous logarithmic term which was found earlier in \cite{Fursaev,Mann,Majumdar,Suneeta,Das,More,Sudipta,Mitra,Page}. We also analyse the Kerr metric which is not spherically symmetric and static, but stationary.

      The organisation of our paper is as follows. In the second section we give a brief review of the radial null geodesic method and its limitations. Section 3 is devoted for the Hamilton-Jacobi method beyond the semiclassical approximation where we derive a general expression for Hawking temperature and then show how one can reproduce the loop corrections to the surface gravity. Also, the Hamilton-Jacobi method in Painleve coordinates is discussed here. In the next two sections the derivations of corrected temperature and entropy for some well known spacetime metrics is given. The well known logarithmic corrections to the area law are reproduced. The final section is for conclusions.

\section{Radial null geodesic method: a brief review}
        In this section we will briefly discuss about the radial null geodesic method \cite{Wilczek} to find the temperature of a black hole using the picture of Hawking radiation as quantum tunneling. Then we will show how one can include the loop corrections due to the back reaction effect in the spacetime. We also point out about the possible limitations of the method.

        This method involves calculating the imaginary part of the action for the (classically forbidden) process of s-wave emission across the horizon which in turn is related to the Boltzmann factor for emission at the Hawking temperature. We consider a general class of static (i.e. invariant under time reversal as well as stationary), spherically symmetric spacetime of the form
\begin{eqnarray}
ds^2 = -f(r)dt^2+\frac{dr^2}{g(r)}+r^2 d\Omega^2
\label{1.01}
\end{eqnarray}
where the horizon $r=r_H$ is given by $f(r_H)=g(r_H)=0$. The coordinate singularity at the horizon is removed by going to Painleve coordinates \cite{Painleve}. Under the transformation
\begin{eqnarray}
dt\rightarrow dt-\sqrt{\frac{1-g(r)}{f(r)g(r)}}dr
\label{1.011}
\end{eqnarray}
the metric (\ref{1.01}) takes the form
\begin{eqnarray}
ds^2=-f(r)dt^2+2f(r)\sqrt{\frac{1-g(r)}{f(r)g(r)}}dtdr+dr^2+r^2d\Omega^2
\label{1.012}
\end{eqnarray}
This metric is stationary but not static. It has a number of interesting features. At any fixed time the spatial geometry is flat and for any fixed radius the boundary geometry is the same as that of the metric (\ref{1.01}).

          The basic idea of this method is to find the radial null geodesics ($ds^2=d\Omega^2=0$) for the metric (\ref{1.012}): 
\begin{eqnarray}
\dot{r}= \sqrt{\frac{f(r)}{g(r)}}(\pm1-\sqrt{1-g(r)})
\label{1.013}
\end{eqnarray}
where $+(-)$ sign gives outgoing (ingoing) null radial geodesics. Using this one has to calculate the imaginary part of the action for a shell of energy $\omega$. In the original work \cite{Wilczek}, the imaginary part of the action is defined as,
\begin{eqnarray}
{\textrm{Im}} S&=& {\textrm{Im}}\int_{r_{\textrm{in}}}^{r_{\textrm{out}}}p_r dr={\textrm{Im}}\int_{r_{\textrm{in}}}^{r_{\textrm{out}}}\int_0^{p_r} dp'_r dr
\nonumber
\\
&=&{\textrm{Im}}\int_{r_{\textrm{in}}}^{r_{\textrm{out}}}\int_0^H\frac{dH'}{\dot r} dr
\label{1.014}
\end{eqnarray}
where in the last step we multiply and divide the integrand by the two sides of Hamilton's equation $\dot r=\frac{dH}{dp_r}|_r$. 

      Near the horizon one can expand $f(r)$ and $g(r)$ about the horizon $r_H$.
\begin{eqnarray}
&&f(r)=f'(r_H)(r-r_H)+{\cal{O}}((r-r_H)^2)
\nonumber
\\
&&g(r)=g'(r_H)(r-r_H)+{\cal{O}}((r-r_H)^2)
\label{1.017}
\end{eqnarray}
Substituting these in (\ref{1.013}) $\dot r$ can be approximately expressed as
\begin{eqnarray}
\dot r\simeq\frac{1}{2}\sqrt{f'(r_H)g'(r_H)}(r-r_H)
\label{1.018}
\end{eqnarray}
Using this in (\ref{1.014}) the imaginary part of the action is calculated. Now taking the tunneling probability as $\Gamma\sim e^{-\frac{2}{\hbar}{\textrm{Im}}S}$ and equating it with the Boltzmann factor $e^{-\frac{\omega}{T}}$ we find the Hawking temperature as
\begin{eqnarray}
T_H=\frac{\omega\hbar}{2{\textrm{Im}}S}=\frac{\hbar\sqrt{f'(r_H)g'(r_H)}}{4\pi}.
\label{1.021}
\end{eqnarray}
It is easy to confirm that for a Schwarzschild black hole the correct result of $T_H=\frac{\hbar}{8\pi M}$ follows.

      Recently, a problem in this approach has been pointed out in \cite{Borun}. Particularly, it has been shown that $2{\textrm{Im}}\int_{r_{\textrm{in}}}^{r_{\textrm{out}}}p_rdr$ is not canonically invariant and thus it is not a proper observable. The object which is canonically invariant is ${\textrm{Im}}\oint p_rdr$ where the closed path goes across the horizon and back. For ordinary tunneling problems where the tunneling amplitude is the same whether one tunnels from left to right or right to left one finds that $\oint p_rdr=2\int_{r_{\textrm{in}}}^{r_{\textrm{out}}}p_rdr$. However for the Painleve coordinates there is only a barrier to a particle coming from inside the horizon to outside. Particles going from outside to inside do not see a barrier and thus the two expressions are not equivalent. Consequently $\oint p_rdr\neq 2\int_{r_{\textrm{in}}}^{r_{\textrm{out}}}p_rdr$ thereby invalidating the analysis leading to (\ref{1.021}). Rather if one uses the invariant definition $\Gamma\sim e^{-\frac{1}{\hbar}{\textrm{Im}}\oint p_rdr}$, the Hawking temperature is found to be twice the original temperature. This ambiguity has been mentioned in more details in \cite{Singleton,Douglas,Pilling,Nakamura,Vanzo,Kerner}. However our analysis will be based on Hamilton-Jacobi method which is free of this factor two discrepancy. This will be shown in the next section.        

               Situations in which there is a back reaction have to be handled separately, as, for instance, discussed 
for the Schwarzschild case \cite{Majhi}. But it is still not clear how to handle other metrics. For example, in the case of Reissner-Nordstrom metric one has to consider the tunneling of a charged particle. In that case the invariant tunneling rate $\Gamma\sim e^{-\frac{1}{\hbar}{\textrm{Im}}\oint p_rdr}$ will change. There are some papers \cite{Wu,Hu} in which this case is analysed. But these are completely based on the semiclassical approximation. No quantum effects have been included. Also, in this method it is necessary to work in Painleve coordinates to avoid the singularity at the horizon. So one is not allowed to perform the calculations in the original metric coordinates, at least in the standard formulation. Furthermore, this method corresponds to radial null geodesics and so it is valid only for massless particles.

\section{Hamilton-Jacobi method beyond the semiclassical approximation}
     We next consider an alternate method for calculating the imaginary part of the action, making use of the Hamilton-Jacobi equation \cite{Paddy,Vanzo,Kerner}, from which we will calculate the Hawking temperature. The analysis goes beyond the semiclassical approximation by including all possible quantum corrections. The semiclassical Hawking temperature is thereby appropriately altered. Equivalent results are obtained by using either the standard Schwarzschild like coordinates or other types, as for instance, the Painleve ones. We discuss both cases in this section.
\subsection{Schwarzschild like coordinate system}
   Let us consider a massless particle {\footnote{Though we consider only the massless particle, this method is valid for massive case also. It has been shown earlier \cite{Paddy} that ultimately the final expressions are same. Therefore for simplicity we consider only the massless case.}} in the spacetime (\ref{1.01}) described by the Klein-Gordon equation
\begin{eqnarray}
-\frac{\hbar^2}{\sqrt{-g}}\partial_\mu[g^{\mu\nu}\sqrt{-g}\partial_{\nu}]\phi = 0.
\label{1.02}
\end{eqnarray}
For radial trajectories only the $(r-t)$ sector of the metric (\ref{1.01}) is important. Therefore under this metric the Klein-Gordon equation reduces to
\begin{eqnarray}
-\frac{1}{\sqrt{f(r)g(r)}}\partial^2_t\phi +\frac{1}{2}\Big(f'(r)\sqrt{\frac{g(r)}{f(r)}}+g'(r)\sqrt{\frac{f(r)}{g(r)}}\Big)\partial_r\phi + \sqrt{f(r)g(r)}\partial_r^2\phi=0.
\label{1.03}
\end{eqnarray}
The semiclassical wave function satisfying the above equation is obtained by making the standard ansatz for $\phi$ which is
\begin{eqnarray}
\phi(r,t)={\textrm{exp}}\Big[-\frac{i}{\hbar}S(r,t)\Big],
\label{1.04}
\end{eqnarray}  
where $S(r,t)$ is a function which will be expanded in powers of $\hbar$. Substituting into the wave equation (\ref{1.03}), we obtain
\begin{eqnarray}
&&\frac{i}{\sqrt{f(r)g(r)}}\Big(\frac{\partial S}{\partial t}\Big)^2 - i\sqrt{f(r)g(r)}\Big(\frac{\partial S}{\partial r}\Big)^2 - \frac{\hbar}{\sqrt{f(r)g(r)}}\frac{\partial^2 S}{\partial t^2} + \hbar \sqrt{f(r)g(r)}\frac{\partial^2 S}{\partial r^2}
\nonumber
\\
&&+ \frac{\hbar}{2}\Big(\frac{\partial f(r)}{\partial r}\sqrt{\frac{g(r)}{f(r)}}+\frac{\partial g(r)}{\partial r}\sqrt{\frac{f(r)}{g(r)}}\Big)\frac{\partial S}{\partial r}=0.
\label{1.05}
\end{eqnarray}
Expanding $S(r,t)$ in a powers of $\hbar$, we find,
\begin{eqnarray}
S(r,t)&=&S_0(r,t)+\hbar S_1(r,t)+\hbar^2 S_2(r,t)+...........
\nonumber
\\
&=&S_0(r,t)+\sum_i \hbar^i S_i(r,t).
\label{1.06}
\end{eqnarray}
where $i=1,2,3,......$. In this expansion the terms from ${\cal{O}}(\hbar)$ onwards are treated as quantum corrections over the semiclassical value $S_0$. Substituting (\ref{1.06}) in (\ref{1.05}) and equating the different powers of $\hbar$ on both sides, we obtain the following set of equations:
\begin{eqnarray}
\hbar^0~:~&&\Big(\frac{\partial S_0}{\partial t}\Big)^2-f(r)g(r) \Big(\frac{\partial S_0}{\partial r}\Big)^2=0,
\nonumber
\\
\hbar^1~:~&&2i\frac{\partial S_0}{\partial t}\frac{\partial S_1}{\partial t}-2if(r)g(r)\frac{\partial S_0}{\partial r}\frac{\partial S_1}{\partial r}-\frac{\partial^2S_0}{\partial t^2}+f(r)g(r)\frac{\partial^2 S_0}{\partial r^2}
\nonumber
\\
&&+\frac{1}{2}\Big(f'(r)g(r)+f(r)g'(r)\Big)\frac{\partial S_0}{\partial r}=0,
\nonumber
\\
\hbar^2~:~&&i\Big(\frac{\partial S_1}{\partial t}\Big)^2+2i\frac{\partial S_0}{\partial t}\frac{\partial S_2}{\partial t}-if(r)g(r)\Big(\frac{\partial S_1}{\partial r}\Big)^2-2if(r)g(r)\frac{\partial S_0}{\partial r}\frac{\partial S_2}{\partial r}-\frac{\partial^2S_1}{\partial t^2}+f(r)g(r)\frac{\partial^2 S_1}{\partial r^2}
\nonumber
\\
&&+\frac{1}{2}\Big(f'(r)g(r)+f(r)g'(r)\Big)\frac{\partial S_1}{\partial r}=0,
\nonumber
\\
.
\nonumber
\\
.
\nonumber
\\
.
\label{15}
\end{eqnarray} 
so on. Now it is interesting to note that any equation in the above set can be simplified by using the equations coming before it. This leads to an identical set of relations,
\begin{eqnarray}
\hbar^0~:~\frac{\partial S_0}{\partial t}=\pm \sqrt{f(r)g(r)}\frac{\partial S_0}{\partial r},
\label{1.08}
\end{eqnarray}
\begin{eqnarray}
\hbar^1~:~&&\frac{\partial S_1}{\partial t}=\pm \sqrt{f(r)g(r)}\frac{\partial S_1}{\partial r},
\nonumber
\\
\hbar^2~:~&&\frac{\partial S_2}{\partial t}=\pm \sqrt{f(r)g(r)}\frac{\partial S_2}{\partial r},
\nonumber
\\
.
\nonumber
\\
.
\nonumber
\\
.
\nonumber
\end{eqnarray} 
and so on; i.e. the functional form of the above set of linear differential equations is same. Therefore the solutions of these equations are not independent and $S_i$'s are proportional to $S_0$. Since $S_0$ has the dimension of $\hbar$ the proportionality constants should have the dimension of inverse of $\hbar^i$. Again in the units $G=c=k_B=1$ the Planck constant $\hbar$ is of the order of square of the Planck Mass $M_P$ and so from dimensional analysis the proportionality constants have the dimension of $M^{-2i}$ where $M$ is the mass of black hole. Specifically, for Schwarzschild type black holes having mass as the only macroscopic parameter, these considerations show that the most general expression for $S$, following from (\ref{1.06}), is given by,
\begin{eqnarray}
S(r,t)&=& S_0(r,t) + \sum_i \beta_i \frac{\hbar^i}{M^{2i}}S_0(r,t)
\nonumber
\\
&=&\Big(1+\sum_i\beta_i\frac{\hbar^i}{M^{2i}}\Big)S_0(r,t).
\label{1.07}
\end{eqnarray}
where $\beta_i$'s are dimensionless constant parameters.

       To obtain a solution for $S(r,t)$ it is therefore enough to solve for $S_0(r,t)$ which satisfies (\ref{1.08}). In fact the standard Hamilton-Jacobi solution determined by this $S_0(r,t)$ is just modified by a prefactor to yield the complete solution for $S(r,t)$. Since the metric (\ref{1.01}) is stationary it has timelike Killing vectors. Thus we will look for solutions of (\ref{1.08}) which behave as 
\begin{eqnarray}
S_0=\omega t + \tilde S_0(r),
\label{2}
\end{eqnarray} 
where $\omega$ is the energy of the particle. Substituting this in (\ref{1.08}) and then integrating we obtain,
\begin{eqnarray}
\tilde{S_0}(r) =  \pm \omega\int_0^r\frac{dr}{\sqrt{f(r)g(r)}}
\label{3}
\end{eqnarray} 
where the limits of the integration are chosen such that the particle goes through the horizon $r=r_H$. The $+ (-)$ sign in front of the integral indicates that the particle is ingoing (outgoing). Using (\ref{2}) and (\ref{3}) in (\ref{1.07}) we obtain
\begin{eqnarray}
S(r,t)= \Big(1+\sum_i\beta_i\frac{\hbar^i}{M^{2i}}\Big)\Big[\omega t  \pm\omega\int_0^r\frac{dr}{\sqrt{f(r)g(r)}}\Big].
\label{1.10}
\end{eqnarray}  
Therefore the ingoing and outgoing solutions of the Klein-Gordon equation (\ref{1.02}) under the back ground metric (\ref{1.01}) is given by exploiting (\ref{1.04}) and (\ref{1.10}),
\begin{eqnarray}
\phi_{{\textrm {in}}}= {\textrm{exp}}\Big[-\frac{i}{\hbar}(1+\sum_i\beta_i\frac{\hbar^i}{M^{2i}})\Big(\omega t  +\omega\int_0^r\frac{dr}{\sqrt{f(r)g(r)}}\Big)\Big]
\label{1.11}
\end{eqnarray} 
and
\begin{eqnarray}
\phi_{{\textrm {out}}}= {\textrm{exp}}\Big[-\frac{i}{\hbar}(1+\sum_i\beta_i\frac{\hbar^i}{M^{2i}})\Big(\omega t  -\omega\int_0^r\frac{dr}{\sqrt{f(r)g(r)}}\Big)\Big].
\label{1.12}
\end{eqnarray} 
Now for the tunneling of a particle across the horizon the nature of the coordinates change. The sign of the metric coefficients in the $(r-t)$ sector is altered. This indicates that `$t$' coordinate has an imaginary part for the crossing of the horizon of the black hole and correspondingly there will be a temporal contribution to the probabilities for the ingoing and outgoing particles. This has similarity with\cite{Akhmedov} where they show for the Schwarzschild metric that two patches across the horizon are connected by a discrete imaginary amount of time.

     The ingoing and outgoing probabilities of the particle are, therefore, given by,
\begin{eqnarray}
P_{{\textrm{in}}}=|\phi_{{\textrm {in}}}|^2= {\textrm{exp}}\Big[\frac{2}{\hbar}(1+\sum_i\beta_i\frac{\hbar^i}{M^{2i}})\Big(\omega{\textrm{Im}}~t +\omega{\textrm{Im}}\int_0^r\frac{dr}{\sqrt{f(r)g(r)}}\Big)\Big]
\label{1.13}
\end{eqnarray}
and
\begin{eqnarray}
P_{{\textrm{out}}}=|\phi_{{\textrm {out}}}|^2= {\textrm{exp}}\Big[\frac{2}{\hbar}(1+\sum_i\beta_i\frac{\hbar^i}{M^{2i}})\Big(\omega{\textrm{Im}}~t -\omega{\textrm{Im}}\int_0^r\frac{dr}{\sqrt{f(r)g(r)}}\Big)\Big]
\label{1.14}
\end{eqnarray}
Now the ingoing probability $P_{\textrm{in}}$ has to be unity in the classical limit (i.e. $\hbar\rightarrow 0$) - when there is no reflection and everything is absorbed - instead of zero or infinity \cite{Partha}.Thus, in the classical limit, (\ref{1.13}) leads to,
\begin{eqnarray}
{\textrm{Im}}~t = -{\textrm{Im}}\int_0^r\frac{dr}{\sqrt{f(r)g(r)}}.
\label{1.16}
\end{eqnarray}
From the above one can easily show that ${\textrm{Im}}~t = -2\pi M$ for the Schwarzschild spacetime which is precisely the imaginary part of the transformation $t\rightarrow t-2i\pi M$ when one connects the two regions across the horizon as shown in \cite{Akhmedov}.  
Therefore the probability of the outgoing particle is
\begin{eqnarray}
P_{{\textrm{out}}}={\textrm{exp}}\Big[-\frac{4}{\hbar}\omega\Big(1+\sum_i\beta_i\frac{\hbar^i}{M^{2i}}\Big){\textrm{Im}}\int_0^r\frac{dr}{\sqrt{f(r)g(r)}}\Big].
\label{1.17}
\end{eqnarray}
Now using the principle of ``detailed balance'' \cite{Paddy,Bibhas}
\begin{eqnarray}
P_{{\textrm{out}}}= {\textrm {exp}}\Big(-\frac{\omega}{T_h}\Big)P_{\textrm{in}}={\textrm{exp}} \Big(-\frac{\omega}{T_h}\Big)
\label{1.18}
\end{eqnarray}
we obtain the temperature of the black hole as
\begin{eqnarray}
T_h &=& \frac{\hbar}{4}\Big[(1+\sum_i\beta_i\frac{\hbar^i}{M^{2i}}){\textrm{Im}}\int_0^r\frac{dr}{\sqrt{f(r)g(r)}}\Big]^{-1}
\nonumber
\\
&=&T_H\Big(1+\sum_i\beta_i\frac{\hbar^i}{M^{2i}}\Big)^{-1}
\label{1.19}
\end{eqnarray}
where 
\begin{eqnarray}
T_H = \frac{\hbar}{4}\Big({\textrm{Im}}\int_0^r\frac{dr}{\sqrt{f(r)g(r)}}\Big)^{-1}
\label{1}
\end{eqnarray}
is the standard semiclassical Hawking temperature of the black hole and other terms are the corrections due to the quantum effect. Using this expression and knowing the metric coefficients $f(r)$ and $g(r)$ one can easily find out the temperature of the corresponding black hole.

      Some comments are now in order. The first point is that (\ref{1}) yields a novel form of the semiclassical Hawking temperature. For instance, it can be used for metrics that need not be spherically symmetric. Later, in section $4.3$, this will be exemplified in the case of the Kerr metric. For a spherically symmetric metric it is possible to show that (\ref{1}) reproduces the familiar form (\ref{1.021}). Inserting the near horizon expansion (\ref{1.017}) in (\ref{1}) and performing the contour integration, (\ref{1.021}) is obtained. Note that this form is the standard Hawking temperature found \cite{Vanzo,Kerner} by the Hamilton-Jacobi method. There is no ambiguity regarding a factor of two in the Hawking temperature as reported in the literature \cite{Vanzo,Kerner}. This issue is completely avoided in the present analysis where the standard expression for the Hawking temperature is reproduced.

       The other point is that the form of the solution (\ref{2}) of (\ref{1.08}) is not unique, since any constant multiple of `$S_0$' can be a solution as well. For that case one can easily see that the final expression (\ref{1.19}) for the temperature still remains unchanged. It is only a matter of rescaling the particle energy `$\omega$'. This shows the uniqueness of the expression (\ref{1.19}) for the Hawking temperature.

   We will now show that various choices of the coefficients $\beta_i$ in (\ref{1.19}) correspond to higher order loop corrections to the surface gravity of the black hole, obtained by including back reaction effects \cite{York,Lousto} or by accounting for the trace anomaly \cite{Fursaev}. To see this note that the standard relation between the surface gravity (${\cal{K}}$) and the Hawking temperature ($T_h$) is  
\begin{eqnarray}
T_h = \frac{{\cal{K}}}{2\pi}.
\label{1.20}
\end{eqnarray}
Hence the modified form of the surface gravity of the black hole following from (\ref{1.19}), is given by
\begin{eqnarray}
{\cal{K}} = {\cal{K}}_0\Big(1+\sum_i\beta_i\frac{\hbar^i}{M^{2i}}\Big)^{-1}.
\label{1.21}
\end{eqnarray}
where ${\cal{K}}_0=2\pi T_H$ is the standard semiclassical surface gravity.

     Now if we choose the dimensionless parameters $\beta_i$'s in terms of a single dimensionless parameter $\alpha$ in the following way,
\begin{eqnarray}
\beta_i = \alpha^i
\label{1.22}
\end{eqnarray}
then the expression within the parenthesis in (\ref{1.21}) is simplified to
\begin{eqnarray}
1+\sum_i\beta_i\frac{\hbar^i}{M^{2i}} &=& 1+\Big(\frac{\alpha\hbar}{M^2}+ \frac{\alpha^2\hbar^2}{M^4}+\frac{\alpha^3\hbar^3}{M^6}+\frac{\alpha^4\hbar^4}{M^8}+........\Big)=(1-\frac{\alpha\hbar}{M^2})^{-1}.
\label{1.23}
\end{eqnarray}
Therefore under this choice of the parameters the surface gravity (\ref{1.21}) simplifies to the following form
\begin{eqnarray}
{\cal{K}}={\cal{K}}_0\Big(1-\frac{\alpha\hbar}{M^2}\Big).
\label{1.27}
\end{eqnarray}
This expression was found earlier \cite{York,Lousto} by considering the one loop back reaction effects in the spacetime. Moreover, the coefficient $\alpha$ is related to the trace anomaly. Using conformal field theory techniques it was shown \cite{Fursaev} that the one loop quantum correction to the surface gravity, for the Schwarzschild black hole, is given by (\ref{1.27}) where,
\begin{eqnarray}
\alpha=-\frac{1}{360\pi}\Big(-N_0-\frac{7}{4}N_{\frac{1}{2}}+13 N_1+\frac{233}{4}N_{\frac{3}{2}}-212N_2\Big)
\label{conformal}
\end{eqnarray}
and $N_s$ denotes the number of fields with spin `$s$'.

       Likewise, the higher order loop corrections in the surface gravity can also be reproduced by this method. The only important part is to choose the expansion coefficients $\beta_i$'s suitably. For instance, if $\beta_i$'s are chosen as below,
\begin{eqnarray}
\beta_i=\sum_{k=0}^\infty (i-k)_{C_k}\alpha^{(i-2k)}\gamma^k
\label{16}
\end{eqnarray}
with $i\geq2k$, then substituting this in (\ref{1.21}) and simplifying, one can show that
\begin{eqnarray}
{\cal{K}}={\cal{K}}_0\Big(1-\frac{\alpha\hbar}{M^2}-\frac{\gamma\hbar^2}{M^4}\Big)
\label{17}
\end{eqnarray}
which is nothing but the corrected form of the surface gravity of a black hole due to two loop back reaction effects in the spacetime. $`\gamma$' is a dimensionless parameter corresponding to the contribution from the second loop. This indicates that one can reproduce the all loop back reaction effects in the spacetime by keeping all the terms in the expansion of $S(r,t)$ (\ref{1.06}) and suitably choosing the expansion coefficients.

\subsection{Painleve coordinate system}
            Here we will discuss the Hamilton-Jacobi method in Painleve coordinates and explicitly show how one can obtain the standard Hawking temperature. As before, consider a massless scalar particle in the spacetime metric (\ref{1.012}) described by the Painleve coordinates. Since the Klein-Gordon equation (\ref{1.02}) is in covariant form, the scalar particle in the background metric (\ref{1.012}) also satisfies (\ref{1.02}). Therefore under this metric the Klein-Gordon equation reduces to
\begin{eqnarray}
-(\frac{g}{f})^{\frac{3}{2}}\partial^2_t\phi+\frac{2g\sqrt{1-g}}{f}\partial_t\partial_r \phi-\frac{gg'}{2f\sqrt{1-g}}\partial_t\phi+g\sqrt{\frac{g}{f}}\partial^2_r\phi+\frac{1}{2}\sqrt{\frac{g}{f}}(3g'-\frac{f'g}{f})\partial_r\phi=0.
\label{18}
\end{eqnarray}
As before, substituting the standard ansatz (\ref{1.04}) for $\phi$ in the above equation, we obtain,
\begin{eqnarray}
&&-(\frac{g}{f})^{\frac{3}{2}}\Big[-\frac{i}{\hbar}\Big(\frac{\partial S}{\partial t}\Big)^2+\frac{\partial^2S}{\partial t^2}\Big] + \frac{2g\sqrt{1-g}}{f}\Big[-\frac{i}{\hbar}\frac{\partial S}{\partial t}\frac{\partial S}{\partial r}+\frac{\partial^2S}{\partial r\partial t}\Big]-\frac{gg'}{2f\sqrt{1-g}}\frac{\partial S}{\partial t}
\nonumber
\\
&&+g\sqrt{\frac{g}{f}}\Big[-\frac{i}{\hbar}\Big(\frac{\partial S}{\partial r}\Big)^2+\frac{\partial^2S}{\partial r^2}\Big]+\frac{1}{2}(3g'-\frac{f'g}{f})\frac{\partial S}{\partial r}=0.
\label{19}
\end{eqnarray}
Neglecting the terms of order $\hbar$ and greater we find to the lowest order,
\begin{eqnarray}
(\frac{g}{f})^{\frac{3}{2}}\Big(\frac{\partial S}{\partial t}\Big)^2-\frac{2g\sqrt{1-g}}{f}\frac{\partial S}{\partial t}\frac{\partial S}{\partial r}-g\sqrt{\frac{g}{f}}\Big(\frac{\partial S}{\partial r}\Big)^2=0.
\label{20}
\end{eqnarray}
It has been stated earlier that the metric (\ref{1.012}) is stationary. Therefore following the same argument as before it has a solution of the form (\ref{2}). Inserting this in (\ref{20}) yields,
\begin{eqnarray}
\frac{d\tilde S_0(r)}{dr}=\omega\sqrt{\frac{1-g(r)}{f(r)g(r)}}\Big(-1\pm\frac{1}{\sqrt{1-g(r)}}\Big)
\label{22}
\end{eqnarray}
Integrating,
\begin{eqnarray}
\tilde S_0(r)=\omega\int_0^r\sqrt{\frac{1-g(r)}{f(r)g(r)}}\Big(-1\pm\frac{1}{\sqrt{1-g(r)}}\Big)dr.
\label{23}
\end{eqnarray}
The $+(-)$ sign in front of the integral indicates that the particle is ingoing (outgoing). Therefore the actions for ingoing and outgoing particles are
\begin{eqnarray}
S_{\textrm {in}}(r,t)=\omega t +\omega\int_0^r\frac{1-\sqrt{1-g}}{\sqrt{fg}}dr
\label{24}
\end{eqnarray}
and
\begin{eqnarray}
S_{\textrm {out}}(r,t)=\omega t -\omega\int_0^r\frac{1+\sqrt{1-g}}{\sqrt{fg}}dr
\label{25}
\end{eqnarray}
Since in the classical limit (i.e. $\hbar\rightarrow 0$) the probability for the ingoing particle ($P_{\textrm{in}}$) has to be unity, $S_{\textrm{in}}$ must be real. Following identical steps employed in deriving (\ref{1.16}) we obtain, starting from (\ref{24}), the analogous condition,
\begin{eqnarray}
{\textrm{Im}}~t=-{\textrm{Im}}\int_0^r\frac{1-\sqrt{1-g}}{\sqrt{fg}}dr
\label{26}
\end{eqnarray}
Substituting this in (\ref{25}) we obtain the action for the outgoing particle:
\begin{eqnarray}
S_{\textrm {out}}(r,t)=\omega{\textrm{Re}}~t -\omega{\textrm{Re}}\int_0^r\frac{1+\sqrt{1-g}}{\sqrt{fg}}dr-2i\omega{\textrm{Im}}\int_0^r\frac{dr}{\sqrt{fg}}
\label{27}
\end{eqnarray}
Therefore the probability for the outgoing particle is
\begin{eqnarray}
P_{\textrm{out}}=|e^{-\frac{i}{\hbar}S_{\textrm{out}}}|^2=e^{-\frac{4}{\hbar}\omega{\textrm{Im}}\int_0^r\frac{dr}{\sqrt{f(r)g(r)}}}
\label{28}
\end{eqnarray}
Now using the principle of ``detailed balance'' (\ref{1.18}) we obtain the same expression (\ref{1}) for the standard Hawking temperature which was calculated in Schwarzschild like coordinates by the Hamilton-Jacobi method. Inclusion of higher order terms is straightforward and leads to the same relation as (\ref{1.19}).

\section{Calculation of Hawking temperature}
     In this section we will consider some standard metrics to show how the semiclassical Hawking temperature can be calculated from (\ref{1}). For the Schwarzschild and Anti-de Sitter Schwarzschild black hole the modified form of the Hawking temperatures due to the corrections beyond semiclassical approximation will be explicitly shown by using (\ref{1.19}).

\subsection{Schwarzschild black hole}
   The spacetime metric is given by
\begin{eqnarray}
ds^2=-(1-\frac{2M}{r})dt^2+(1-\frac{2M}{r})^{-1}dr^2+r^2d\Omega^2.
\label{3.01}
\end{eqnarray}
So the metric coefficients are
\begin{eqnarray}
f(r)=g(r)=(1-\frac{r_H}{r});\,\,\,r_H = 2M.
\label{3.02}
\end{eqnarray}
Since this metric is spherically symmetric we use the formula (\ref{1.021}) to compute the semiclassical Hawking temperature. This is found to be, 
\begin{eqnarray}
T_H=\frac{\hbar}{4\pi r_H}=\frac{\hbar}{8\pi M}.
\label{3.05}
\end{eqnarray}
which is the standard expression.
Now using (\ref{1.19}) it is easy to write the corrected Hawking temperature:
\begin{eqnarray}
T_h=\frac{\hbar}{8\pi M}\Big(1+\sum_i\beta_i\frac{\hbar}{M^{2i}}\Big)^{-1}.
\label{3.06}
\end{eqnarray}
In particular, for the choice (\ref{1.22}), this yields,
\begin{eqnarray}
T_h=\frac{\hbar}{8\pi M}\Big(1-\frac{\alpha\hbar}{M^{2}}\Big)
\label{3.08}
\end{eqnarray}
which is the modified form of the Hawking temperature. Such a structure was obtained earlier \cite{Majhi} in radial null geodesic approach by explicitly taking into account the one loop back reaction effect. Also, as stated earlier, such a form follows from conformal field theory techniques \cite{Fursaev} where $\alpha$ is given by (\ref{conformal}).

\subsection{Anti-de Sitter Schwarzschild black hole}
   The AdS-Schwarzschild metric is given by
\begin{eqnarray}
ds^2=-\Big(1-\frac{2M}{r}+\frac{r^2}{l^2}\Big)dt^2+\Big(1-\frac{2M}{r}+\frac{r^2}{l^2}\Big)^{-1}dr^2+r^2d\Omega^2.
\label{4}
\end{eqnarray}
Here,
\begin{eqnarray}
f(r)=g(r)=\Big(1-\frac{2M}{r}+\frac{r^2}{l^2}\Big).
\label{conf1}
\end{eqnarray}
Since the metric is spherically symmetric we employ similar steps as before to obtain the semiclassical Hawking temperature,
\begin{eqnarray}
T_H=\frac{\hbar}{4\pi}\frac{3r_+^2+l^2}{l^2r_+}.
\label{6}
\end{eqnarray}
Also, by (\ref{1.19}), the corrected form of the Hawking temperature due to quantum effects, is
\begin{eqnarray}
T_h=\frac{\hbar}{4\pi}\frac{3r_+^2+l^2}{l^2r_+}\Big(1+\sum_i\beta_i\frac{\hbar^i}{M^{2i}}\Big)^{-1}.
\label{7}
\end{eqnarray}
Once again, for the choice (\ref{1.22}), this yields,
\begin{eqnarray}
T_h= \frac{\hbar}{4\pi}\frac{3r_+^2+l^2}{l^2r_+}\Big(1-\frac{\alpha\hbar}{M^2}\Big)
\label{cong12}
\end{eqnarray}
which reproduces the corrected Hawking temperature by including the one loop back reaction effect \cite{Lousto}.


\subsection{Kerr black hole}
     This example provides a nontrivial application of our formula (\ref{1}) for computing the semiclassical Hawking temperature. Here the metric is not spherically symmetric, invalidating the use of (\ref{1.021}).

     In Boyer-Linquist coordinates the form of the Kerr metric is given by
\begin{eqnarray}
ds^2&=&-\Big(1-\frac{2Mr}{\rho^2}\Big)dt^2-\frac{2Mar~{\textrm {sin}}^2\theta}{\rho^2}(dt d\phi+d\phi dt)
\nonumber
\\
&+&\frac{\rho^2}{\Delta}dr^2+\rho^2d\theta^2+\frac{{\textrm {sin}}^2\theta}{\rho^2}~\Big[(r^2+a^2)^2-a^2\Delta~{\textrm {sin}}^2\theta\Big]d\phi^2
\label{3.21}
\end{eqnarray}
where
\begin{eqnarray}
\Delta(r)&=&r^2-2Mr+a^2;\,\,\,\rho^2(r,\theta)=r^2+a^2~{\textrm{cos}}^2\theta
\nonumber
\\
a&=&\frac{J}{M}
\label{3.22}
\end{eqnarray}
and $J$ is the Komar angular momentum. We have chosen the coordinates for Kerr metric such that the event horizons occur at those fixed values of $r$ for which $g^{rr}=\frac{\Delta}{\rho^2}=0$. Therefore the event horizons are
\begin{eqnarray}
r_\pm = M\pm\sqrt{M^2-a^2}.
\label{3.23}
\end{eqnarray}
This metric is not spherically symmetric and static but stationary. So it must have timelike Killing vectors. Although in our general formulation we consider only the static, spherically symmetric metrics, it is still possible to apply this methodology for such a metric. The point is that for radial trajectories, the Kerr metric simplifies to the following form
\begin{eqnarray}
ds^2=-\Big(\frac{r^2+a^2-2Mr}{r^2+a^2}\Big)dt^2+\Big(\frac{r^2+a^2}{r^2+a^2-2Mr}\Big)dr^2
\label{3.24}
\end{eqnarray}
where, for simplicity, we have taken $\theta=0$ (i.e. particle is going along $z$-axis). This is exactly the form of the $(r-t)$ sector of the metric (\ref{1.01}). Since in our formalism only the $(r-t)$ sector is important, our results are applicable here. In particular if the metric has no terms like $(dr dt)$ then we can apply (\ref{1}) to find the semiclassical Hawking temperature. Here,
\begin{eqnarray}
f(r)=g(r)=\Big(\frac{r^2+a^2-2Mr}{r^2+a^2}\Big)
\label{3.25}
\end{eqnarray}
Substituting these in (\ref{1}) we obtain,
\begin{eqnarray}
T_H=\frac{\hbar}{4}\Big({\textrm{Im}}\int\frac{r^2+a^2}{(r-r_+)(r-r_-)}\Big)^{-1}.
\label{conf5}
\end{eqnarray}
The integrand has simple poles at $r=r_+$ and $r=r_-$. Since we are interested only with the event horizon at $r=r_+$, we choose the contour as a small half-loop going around this pole from left to right. Integrating, we obtain the value of the semiclassical Hawking temperature as
\begin{eqnarray}
T_H=\frac{\hbar}{4\pi}\frac{r_+ - r_-}{r_+^2+a^2}.
\label{3.26}
\end{eqnarray}
This agrees with results quoted in the literature \cite{Carrol}.

\section{Bekenstein-Hawking entropy and area law}
         The semiclassical Bekenstein-Hawking area law \cite{Bekenstein,Bardeen,Hawking2} is given by 
\begin{eqnarray}
S_{\textrm{BH}}=\frac{A}{4\hbar}
\label{11}
\end{eqnarray}
where `$A$' is the area of the horizon. This is altered when quantum effects come into play. Here, using the modified form of the temperature for Schwarzschild and AdS-Schwarzschild black hole derived in the previous section, we will explicitly show the corrections to the Bekenstein-Hawking area law with the help of the second law of thermodynamics. This law of black hole thermodynamics which expresses the conservation of energy by relating the change in the black hole mass $M$ to the changes in its entropy $S_{\textrm{bh}}$, angular momentum $J$ and electric charge $Q$, is given by 
\begin{eqnarray}
dM = T_h dS_{\textrm{bh}} + \Phi dQ + \Omega dJ
\label{1.28}
\end{eqnarray}
where the angular velocity $\Omega$ and the electrostatic potential $\Phi = \frac{\partial M}{\partial Q}$ are constant over the event horizon of any stationary black hole. From this conservation law the entropy is computed as,
\begin{eqnarray}
S_{\textrm{bh}}=\int\frac{1}{T_h}(dM-\Phi dQ-\Omega dJ)
\label{conf2}
\end{eqnarray}

\subsection{Schwarzschild black hole}
   It has no charge and spin. Hence (\ref{conf2}) simplifies to
\begin{eqnarray}
S_{\textrm{bh}}=\int\frac{dM}{T_h}
\label{conf3}
\end{eqnarray}
Substituting the value of temperature from (\ref{3.06}) in (\ref{conf3}) we obtain,
\begin{eqnarray}
S_{\textrm{bh}}&=& \frac{4\pi M^2}{\hbar}+8\pi \beta_1 \ln M - \frac{4\pi \hbar \beta_2}{M^2}+{\textrm{higher order terms in $\hbar$}}
\nonumber
\\
&=&\frac{\pi r_H^2}{\hbar}+8\pi\beta_1\ln r_H -\frac{16\pi\hbar\beta_2}{r_H^2}+{\textrm{higher order terms in $\hbar$}}
\label{12}
\end{eqnarray}
The area of the event horizon is
\begin{eqnarray}
A=4\pi r_H^2
\label{13}
\end{eqnarray}
so that, 
\begin{eqnarray}
S_{\textrm{bh}}&=&\frac{A}{4\hbar}+4\pi\beta_1\ln A+\frac{64\pi^2\hbar\beta_2}{A}+......................
\label{conf4}
\end{eqnarray}
It is noted that the first term is the usual semiclassical area law (\ref{11}). The other terms are the quantum corrections. Now it is possible to express the quantum corrections in terms of $S_{\textrm{BH}}$ by eliminating $A$:
\begin{eqnarray}
S_{\textrm{bh}}=S_{\textrm{BH}}+4\pi\beta_1\ln S_{\textrm{BH}}+\frac{16\pi^2\beta_2}{S_{\textrm{BH}}}+.........
\label{4.06}
\end{eqnarray}
Interestingly the leading order correction is logarithmic in $A$ or $S_{\textrm{BH}}$ which was found earlier in \cite{Fursaev,Mann} by field theory calculations and later in \cite{Majumdar,Mitra} with $\beta_1=-\frac{1}{8\pi}$ by quantum geometry method. The higher order corrections involve inverse powers of $A$ or $S_{\textrm{BH}}$.

\subsection{Anti-de Sitter Schwarzschild black hole}
      Since this black hole also does not have charge and spin, the appropriate equation defining the entropy is given by (\ref{conf3}). The first point to observe is that the event horizon $r=r_+$ is defined by
\begin{eqnarray}
\Big(1-\frac{2M}{r_+}+\frac{r_+^2}{l^2}\Big)=0.
\label{5}
\end{eqnarray}
This shows that it is possible to interpret $M$ as a function of $r_+$;$M=M(r_+)$. Hence $dM=\frac{\partial M}{\partial r_+}dr_+$. Calculating $\frac{\partial M}{\partial r_+}$ from (\ref{5}) we obtain
\begin{eqnarray}
dM=\frac{3r_+^2+l^2}{2l^2}dr_+.
\label{9}
\end{eqnarray}
Substituting this and (\ref{7}) in (\ref{conf3}) and integrating, we obtain the corrected Bekenstein-Hawking entropy as
\begin{eqnarray}
S_{\textrm{bh}}&=&\frac{\pi r_+^2}{\hbar}+8\pi\beta_1\ln r_+ + .........
\label{14}
\end{eqnarray}
Here also the area $A$ satisfies (\ref{13}). Re-expressing (\ref{14}) in terms of the semiclassical expression $S_{\textrm{BH}}$ (\ref{11}), we obtain,
\begin{eqnarray}
S_{\textrm{bh}}=S_{\textrm{BH}}+4\pi\beta_1\ln S_{\textrm{BH}}+.............
\label{10}
\end{eqnarray}
The leading order correction is again logarithmic as found earlier in \cite{Das,More} by a statistical method with $\beta_1=-\frac{1}{4\pi}$.



\section{Conclusions}
      We have given a general expression (\ref{1}) for the semiclassical Hawking temperature by the Hamilton-Jacobi method which corresponds directly to the picture that visualises the source of radiation as tunneling. For the particular case of a spherically symmetric metric, our expression reduces to the standard form (\ref{1.021}). Going beyond the semiclassical approximation, the one particle action is computed by including all higher order corrections. From this action, the modified Hawking temperature for the Schwarzschild and AdS-Schwarzschild black holes are given. The Hamilton-Jacobi method is also studied in other coordinates like Painleve. Here also the Hawking temperature is explicitly calculated which exactly matches with that evaluated in Schwarzschild like coordinates. In all these cases there is no ambiguity regarding a factor of two in the temperature, as reported in \cite{Singleton,Pilling,Vanzo,Kerner}.

       In this paper, the factor of two problem in the Hawking temperature has been taken care of by considering the contribution from the imaginary part of the temporal coordinate since it changes its nature across the horizon. Also, this method is free of the rather ad hoc way of introducing an integration constant, as reported in \cite{Partha}. Our approach, on the other hand, is similar to in spirit \cite{Akhmedov} where it has been shown that `$t$' changes by an imaginary discrete amount across the horizon. Indeed, the explicit expression for this change, in the case of Schwarzschild metric, calculated from our general formula (\ref{1.16}) agrees with the findings of \cite{Akhmedov}.

      The other significant point of this paper is that for an appropriate choice of the dimensionless parameters appearing in the single particle action, it is possible to reproduce the one loop results obtained by including back reaction effects \cite{York,Lousto} or those based on conformal field theory techniques \cite{Fursaev}. Apart from the one loop case, we have also discussed the nature of two loop corrections to the surface gravity of the black hole. By using a law of black hole mechanics, the corrections to the Bekenstein-Hawking area law are given for Schwarzschild and AdS-Schwarzschild black holes. Interestingly, the leading order correction is a logarithmic function of the horizon area $A$ or the semiclassical entropy $S_{\textrm{BH}}$, as reported earlier in \cite{Fursaev,Mann,Majumdar,Sudipta,Mitra,Suneeta,Das,More,Page}.

       An important part of the analysis is that this is done in four dimensions and the expansion coefficients, apart from some undetermined parameters, are identified by a dimensional analysis. Also, explicit higher order computations were presented for Schwarzschild type black holes that have mass as the only macroscopic parameter. As was discussed, this simplified the dimensional analysis allowing for a compact expression for the modified one particle action. It remains an open issue to extend this analysis to include higher order corrections for other black hole geometries.

\end{document}